# The new microscopic Vavilov-Cherenkov radiation theory

S.G. Chefranov

A.M. Obukhov Institute of Atmospheric Physics RAS, Moscow, Russia, schefranov@mail.ru

Abstract

It is proposed the new microscopic theory of Vavilov-Cherenkov radiation (VCR), emitted directly by medium in non-equilibrium state, arising due to the interaction of medium with a sufficiently fast charged particle. Contrary to the macroscopic VCR theory of Tamm-Frank and quantum VCR theory of Ginzburg, we establish the new VCR parametric resonance mechanism and the new threshold of VCR effect, which is better corresponding to the VCR observations. We show that counting of the, known from Abraham's electrodynamics, force density $F_A = \frac{1}{2} rot[PE]$ (**P** is the polarization vector of the medium in locally non-equilibrium anisotropic state, arising due to the non-stationary electric field **E** of particle, moving with the constant speed **v**) defines the possibility of parametric resonance excitation of the transversal to **E** polarization waves **P**. We get that the condition of exponential growth with time of the amplitude of the wave **P**, providing the realization of VCR effect, is $|v| > v_{th} = \frac{c}{n_*}$, where c is the light speed in vacuum, $n_*(n) > n > 1$, and n is the refraction index of the isotropic medium in the equilibrium state.

# The new microscopic Vavilov-Cherenkov radiation theory
## S.G. Chefranov

### § 1. Introduction

It is known that Vavilov-Cherenkov radiation (VCR), emitted by got out from equilibrium medium, has character continuous spectrum of sharply asymmetric radiation cone that is, more over, partially polarized [1-3]. Corresponding degree of the VCR polarization does not depend on the medium parameters (its viscosity, e.g.). The direction of paramount oscillations of electric vector of the VCR field, in all the cases of conducted experiments, is near coincides with the direction of the main primary beam of particles, initiating VCR emitting by the medium itself [2, 3].

In spite of nearly century-old history of VCR effect discovery, up to now, however, there is no its microscopic theory, which defining threshold mechanism of non-stationary emitting of VCR by the very medium and not by charged particles directly, as in the case of accelerated particles [2, 3].

Actually, known VCR theory of Tamm-Frank [2] is purely macroscopic and describes not self non-stationary mechanism of VCR emitting by the medium, but only already established stationary coherent VCR field in the medium, which always returned after VCR emitting to its stationary isotropic equilibrium state (see also [4-6]). In [2], it is just noted: «From the point of view of microscopic theory radiation under consideration is not emitted directly by an electron, but is caused by coherent oscillations of the medium molecules, excited by the electron. We, however, here do not enter the microscopic consideration of the problem ». In [3], it is also pointed that«… we deal actually with the radiation emitted by medium under the influence of the field of moving in it particle ». In the quantum VCR theory of Ginzburg [7] (see also references given in [7] and [8]), as in [2], it is not counted explicitly that VCR emitter is the medium itself. VCR theories [2] and [7] also have the same values of "threshold" for VCR speed of a charged particle $V_{th_v} = \frac{c}{n}$ (c is the light speed in vacuum, n is the refraction index of the medium for n>1) are obtained. As shown in [4,5], this value of $V_{th_0}$ corresponds, however, not to the VCR threshold and boundary of the observed VCR cone, but only to the position of the interferential maximum of the already established coherent VCR field in medium.

The both VCR theories [2, 7] are actually based on the Minkowski's electrodynamics theory, but not on the alternative to it Abraham's theory. At the end of 1960-s, however, it was very Abraham's theory that gained common recognition after direct experimental confirmation of its conclusions (look about it e.g. in [7]). In this relation, in [9] limits of the formal applicability of Minkowski's theory are established which correspond only to the cases of description of stabilized stationary processes in the equilibrium medium with participation of electromagnetic fields.

Hence, for building of microscopic theory of threshold emitting of VCR by non-equilibrium medium, definitely, it is necessary to use very Abraham's electrodynamics, but not Minkowski's one.

In [2], it is noted that using of energy-momentum tensor suggested in Abraham's theory results in acting of additional force density $F_A = \frac{1}{2} rot[PE]$ on the particles of uniform media in the rest. Here **E** is intensity of the external electric field affecting on the medium particles, and **P** is the polarization vector in the medium. For the equilibrium state of isotropic medium characterized by the refraction factor n, value $F_A = 0$, that is noted in [2]. However, it may be not so and $F_A \neq 0$ in the case when electric field **E** is created in the medium by a particle moving in it with speed **v**. Actually, in the medium in that case, there is distinguished direction (defined by the direction of vector **v**) and it may be locally got out from the isotropic state due to the acting of non-stationary field **E** on the medium particles. As a result, value **F**$_A$ in that case is not equal to zero.

In the present work, we show that for defining threshold microscopic mechanism of VCR emitting by medium, it is important counting of the pointed addition to the force density in the form **F**$_A$. We found that VCR emitting may be related with parametric resonance excitation in the medium of linear polarization waves **P**, transversal to the direction of vector **E**, i.e. waves with (**P**, **E**) = 0.

We defined new value of threshold speed $V_{th} = c/n_*$ ($n_* > n > 1$ and also $n_* > 1$ for n<1[4,5]), which under certain conditions coincides with inferred from new quantum VCR theory [4-6], also based on Abraham's theory of electromagnetic field in the medium.

## § 2. Microscopic VCR theory

1. In Abraham's representation energy-momentum tensor of electromagnetic field in the medium at rest is defined by the following representation of values of three-dimensional tension tensor $\sigma^A_{\alpha\beta}(\alpha,\beta=1,2,3)$ and moment density $g^A_\alpha$ [10, 11]:

$$\sigma^A_{\alpha\beta} = \frac{1}{8\pi}\left[E_\alpha D_\beta + E_\beta D_\alpha + H_\alpha B_\beta + H_\beta B_\alpha - \delta_{\alpha\beta}((ED)+(HB))\right], \quad (1)$$

$$g^A_\alpha = \frac{1}{4\pi c}[EH], \quad (2)$$

where $\mathbf{D} = \mathbf{E} + 4\pi\mathbf{P}$, $\mathbf{E}$ is the vector of electric field intensity, $\mathbf{H}$ is the vector of magnetic field intensity(further, everywhere, we assume $\mathbf{B} = \mathbf{H}$), and $\mathbf{P}$ is the medium polarization vector. For (1), (2), we can define the following force density affecting on a charged particle (having charge density $g_e$) [7, 10, 11]:

$$f^A_\alpha = \frac{\partial \sigma^A_{\alpha\beta}}{\partial x_\beta} - \frac{\partial g^A_\alpha}{\partial t} \quad (3)$$

From (3) and (1), (2), one gets

$$\mathbf{f}^A = g_e\mathbf{E} + \frac{1}{c}\left[(\gamma+\dot{\mathbf{P}})\mathbf{H}\right] + \frac{1}{2}rot[\mathbf{PE}] + \frac{1}{c}\frac{\partial}{\partial t}[\mathbf{PH}] + \mathbf{f}^A_0 \quad (4)$$

where $\gamma$ is the current density, dote above $\mathbf{P}$ denotes partial derivative on time, i.e. operator $\frac{\partial}{\partial t}$, and $f^A_{\alpha 0} = (P_\beta \frac{\partial E_\beta}{\partial x_\alpha} - E_\beta \frac{\partial P_\beta}{\partial x_\alpha})/2$. In (4), the third term corresponds to noted above force density $F_A$, that is absent in Minkowski's theory and is character for Abraham's theory.

Using (4), we may consider the following equation defining dependence on $\mathbf{f}^A$ for amplitude of disturbances of linear polarization waves $\mathbf{P}$ in non-equilibrium medium when $\mathbf{E}(t)\neq 0$:

$$\frac{\partial^2 P_\alpha}{\partial t^2} + 2\delta\frac{\partial P_\alpha}{\partial t} - \frac{c^2}{n^2}\Delta P_\alpha = \frac{e}{m}f^A_\alpha, \quad (5)$$

where $\Delta$ is Laplace operator, c is light speed in vacuum, n is refraction index of the medium in the equilibrium state, $\delta^{-1}$ is character time of relaxation of waves **P** for zero-valued right-hand side of (5); e and m are charge and mass of an electron.

2. Let us use (5) in the case when in the medium or in its proximity, an electron moves with constant speed **v**. Assume that the medium does not affect on the electron motion (or assume that the medium's impact is balanced by some other force supporting constancy of the electron speed). For such motion, the electron creates in the medium distinguished direction due its own field that coincides with its field in vacuum and in the Fourier representation has the following form [12]:

$$E_k = \frac{i4\pi e\left(-k + \frac{(kv)v}{c^2}\right)e^{-i(kv)t}}{\left(k^2 - \frac{(kv)^2}{c^2}\right)}, \tag{6}$$

$$H_k = \frac{i4\pi e}{c}[kv]\frac{e^{-i(kv)t}}{\left(k^2 - \frac{(kv)^2}{c^2}\right)}, \tag{7}$$

where $E_k$ and $H_k$ are Fourier representations of electric and magnetic fields of the electron.

When considering (4) – (7), let us limit ourselves, for simplicity, by investigation of evolution in time only orthogonal to **E** component of polarization vector **P**, for which (**EP**) = 0 and |**P**| =P. From (5), in that case (assuming negligible impact of magnetic field on evolution of value P), one gets equation

$$\frac{\partial^2 P}{\partial t^2} + 2\delta\frac{\partial P}{\partial t} - \frac{c^2}{n^2}\Delta P = \frac{e}{2m}P\,divE \tag{8}$$

For (**kv**)≠0 and (**EP**)=0, direction of vector **P** may have character for VCR major direction corresponding to the particle speed vector **v**. Using Fourier representation $P_k$ for P, from (8) and (6), we get equation

$$\frac{\partial^2 P_k}{\partial t^2} + 2\delta\frac{\partial P_k}{\partial t} + \omega_0^2 P_k = \frac{2\pi e^2}{m}e^{-i(kv)t}P(\mathbf{v}t,t) \tag{9}$$

where $\omega_0^2 = c^2 k^2 / n^2$. In the right-hand side of (9), there is function P, depending on $P_k$ only in the integral form since $P(\mathbf{v}t;t) = \frac{1}{(2\pi)^3} \int d^3 q e^{i(qv)t} P_q$.

Let us seek for solution of (9), using representation

$$P_k = \frac{B(t)\pi^{3/2} e^{-\frac{k^2}{4\alpha^2}}}{\alpha^3}, \quad (10)$$

that correspond to the form $P(\mathbf{x};t) = B(t) e^{-\alpha^2 x^2}$ with unknown function of time B(t). The function B(t) defines evolution of the polarization wave small disturbance amplitudes under action of the electron field (6).

From (9) and (10), for polarization wave amplitude B(t), one gets the equation

$$\frac{d^2 B}{dt^2} + 2\delta \frac{dB}{dt} + B\omega_0^2 (1 - \varepsilon \exp(-i\Omega t - \alpha^2 v^2 t^2)) = 0, \quad (11)$$

where $\Omega = k v \cos\theta$, $\varepsilon = \frac{\Omega_e^2 \alpha^3 \exp(k^2/4\alpha^2)}{2\pi^{3/2} \omega_0^2 N}$, $\Omega_e^2 = \frac{4\pi e^2 N}{m}$, N is the number of electrons in the unity volume of medium.

In Appendix, based on application to (11) of the Krylov-Bogolubov method in the frequency parametric resonance region, we show that parametric resonance excitation of the polarization wave P (i.e. exponential increase with time of the amplitude of VCR) may take place (e.g. in the limit of $\varepsilon \ll 1$ and $k/\alpha \ll 1$) under condition (see (A.12)):

$$v > v_{th} = c/n_*, \quad (12)$$

$$n_* = n(1 + \frac{8\Omega_e^2 \alpha^2 k}{\pi^2 \omega_0^2 N}), \quad (13)$$

For n>1 in [4, 5], on the base of the use of Abraham's representation of photon momentum in the medium (when $P_A = \frac{E}{nc}$, E is the photon energy in the medium), it is obtained that $v_{th} = \frac{c}{n(1+\frac{\sqrt{n^2-1}}{n})}$. Here presence of the ratio $\frac{\sqrt{n^2-1}}{n}$ is related with counting in [4, 5] of the finite and real photon mass in the medium $m_{ph}$ (since $m_{ph} = \frac{E\sqrt{n^2-1}}{c^2 n}$ for n>1). From comparison with (13) one can get for $m_{ph}$ the following representation

$$m_{ph} = \frac{8E\Omega_e^2 \alpha^2 k}{c^2 \pi^2 \omega_0^2 N} \quad (14)$$

Hence, on the base of microscopic VCR theory, it is established relation with the medium parameters for introduced in [4,5] value of effective photon mass obtained by it in the medium and also parametrical resonance mechanism of threshold VCR emitting by the medium got out from the equilibrium state under conditions (12),(13). In (12), (13), all values correspond to the medium parameters characterizing its equilibrium isotropic state.

In (12), (13), value n* differs little from n in the considered limit of small values $\frac{k}{\alpha} \ll 1$, when n>1. For finite values of $\frac{k}{\alpha}$, value $v_{th}$ may already non-monotonically depend on that parameter. In that case, it is possible existence of the minimal value $v_{th\ min}$ for some definite $\left(\frac{k}{\alpha}\right)_{min}$. In the result, it is found out that realization of VCR effect manifests itself most significantly under this very relationship of the character spatial scales. Besides, if the corresponding value $v_{thmin}$ does not exceed speed of light in vacuum c, that is possible also for n<1, then it defines the threshold of VCR realization for n<1, as it is stated in [4,5].

**Appendix**

1. Let in (11), amplitude B(t) has the form[13-15]:

$$B(t) = a(t)\cos(\frac{p}{q}\Omega t + \varphi(t)), \quad (A.1)$$

$$\frac{dB}{dt} = -a(t)\frac{p}{q}\Omega \sin(\frac{p}{q}\Omega t + \varphi(t)), \quad (A.2)$$

where p and q are integers.

In (A.1) and (A.2), two new variables a(t) and $\varphi(t)$ are introduced of the action-angle type. It is not difficult to get for them two evolutionary equations using compatibility conditions (A.1) and (A.2), and also result of substitution in (11) of these representations of B and its derivative in time. Let us apply to so far obtained equations for variables a(t) and $\varphi(t)$, averaging method [13] in the region of parametrical frequency resonance. In the result, one gets (under condition $\left|1 - \frac{p\Omega}{q\omega_0}\right| = |\beta_0| < \varepsilon << 1$, see [13-15]) the system:

$$\frac{db}{dt} = -\varepsilon b \omega_0 ((I_+ + I_-)\sin 2\varphi + i(I_+ - I_-)\cos 2\varphi)/2, \quad (A.3)$$

$$\frac{d\varphi}{dt} = \omega_0(1 - \frac{p^2\Omega^2}{q^2\omega_0^2}) - \varepsilon\omega_0(2I_0 + (I_+ + I_-)\cos 2\varphi - i(I_+ - I_-)\sin 2\varphi)/2, \quad (A.4)$$

a(t)=b(t)exp(-$\delta$ t), (A.5)

$$I_\pm = \int_{-1/2}^{1/2} d\tau \exp(-\alpha^2 \upsilon^2 T^2 \tau^2)\cos(\Omega T\tau(1\pm 2\frac{p}{q})), I_0 = \int_{-1/2}^{1/2} d\tau \exp(-\alpha^2 \upsilon^2 T^2 \tau^2)\cos(\Omega T\tau) \quad (A.6)$$

Integrals in (A.6) correspond to averaging over the period T=2$\pi$q/p$\Omega$, and system (A.3), (A.4) has the following invariant (non adiabatic invariant since in the range of parametrical frequency resonance as known invariance of adiabatic invariant of action is not held [14,15]):

$$J = b^2 \frac{d\varphi}{dt} \quad (A.7)$$

Where derivative over time of the angular variable in (A.7) is defined by the expression of the right hand side of (A.4). Existence of invariant (A.7) allows to solving the system(A.3,A4) in quadratures.

In the case of zero value of invariant (A.7), it is possible to get explicit dependence on time for amplitude b(t)=exp($\lambda t$)b(0). It is not difficult to get for real part of the exponent the following expression

$$\text{Re}\,\lambda = \pm \varepsilon \omega_0 \sqrt{I_+ I_- - (I_0 - \frac{\beta_0}{\varepsilon})^2} / 2 \,. \qquad (A.8)$$

Hence, necessary condition of exponential instability of polarization waves due to parametric resonance corresponds to the positive real part of exponent in (A.8). This in its turn is possible only in the case of positive of under-root expression in (A.9), i.e. under condition

$$I_+ I_- > \left(I_0 - \frac{\vartheta_0}{\varepsilon}\right)^2 \qquad (A.9)$$

Obviously, in the limit of $\alpha \upsilon / \Omega \ll 1$, inequality (A.9) is not met because $I_+ I_- < 0$.

From the other side, in the opposite limit $k/\alpha \ll 1$, from (A.8), we get that the value of frequency difference $\beta_0$ must meet more strict condition (than pointed above inequality $|\beta_0| < \varepsilon$) of the form $\beta_0 < \varepsilon \dfrac{\sqrt{\pi k}\cos\theta}{\alpha \pi \frac{P}{q}}$.

Vice versa, bounding from below on the value $\beta_0$ emerges already from the condition of exponential parametric instability of linear polarization waves when it is required meeting of inequality $\text{Re}\,\lambda > \delta$.

In the result, $\beta_0$ must meet the following inequality

$$\frac{4\delta^2 \pi}{c\omega_0^2 \frac{\sqrt{\pi k}\cos\theta}{\alpha \frac{P}{q}}} < \beta_0 < \varepsilon \sqrt{\pi} \frac{k}{\alpha} \frac{\cos\theta}{\pi \frac{P}{q}}$$

(A.10)

Inequality (A.10) can be satisfied only under condition

$$\varepsilon \frac{k \cos\theta}{\alpha \sqrt{\pi \frac{P}{q}}} > \frac{2\delta}{\omega_0}$$

From the right hand side inequality in (A.10), (since $\beta_0 = 1 - \frac{p^2 v^2 n^2 \cos^2\theta}{q^2 c^2}$) a condition defining anisotropy of VCR follows:

$$\cos\theta > -\frac{\varepsilon k c^2}{2\alpha\sqrt{\pi}\left(\frac{P}{q}\right)^3 v^2 h^2} + \left[\frac{\varepsilon^2 k^2 c^4}{4\alpha^2 \pi \left(\frac{P}{q}\right)^6 v^4 h^4} + \frac{c^2}{\frac{P^2}{q^2} v^2 h^2}\right]^{1/2}$$

(A.11)

Neglecting the term $O(\varepsilon^2)$ in (A.11), we get from (A.11) (taking into account $\cos\theta < 1$) the following bound on value $v > v_{th}$, which is necessary for realization of VCR effect in the form of parametric resonant excitation of polarization waves:

$$v > v_{th} = \frac{c}{n*_0},$$

(A.12)

$$n*_0 = \frac{P}{q} n \left(1 + \frac{4\varepsilon k}{\sqrt{\pi \alpha \frac{P}{q}}} + \cdots\right),$$

(A.13)

From (A.13), it follows that the value of threshold velocity decreases with growth of $\frac{P}{q}$. However, meanwhile becomes less also and power of exponential growth with time of amplitude polarization wave disturbances. For the main de-multiplicative resonance with $\frac{P}{q} = \frac{1}{2}$, inversely, for the maximal high velocity of exponential growth of polarization disturbances, it is found that value of threshold velocity $v_{th}$ in (A.12) is relatively large. Hence, optimal for VCR effect realization in the form of parametric polarization waves excitation is the following usually also well observable resonance with $\frac{P}{q} = 1$. For $\frac{P}{q} = 1$, threshold velocity $v_{th}$ in (A.12), (A.13) coincides with given in the main text body representation (12), (13).

In (12) and (A.12), actually $n_{*0}$ only slightly exceeds value n, because these representations for $v_{th}$ are obtained in the limit of small $\frac{k}{\alpha} \ll 1$ for $\varepsilon \ll 1$. For finite values $\frac{k}{\alpha}$, value $v_{th}$ may already non-monotonically depend on this parameter and reach its minimal value for some particular value $\frac{k}{\alpha}$.